\documentclass[]{aipproc}
\layoutstyle{6x9}

\def\gev{{\rm GeV}}
\def\mev{{\rm MeV}}

\def\OMIT#1{{}}

\def\qcut{q_{\rm cut}}
\def\mxcut{m_{\rm cut}}

\def\gcut{G(\qcut^2,\mxcut)}

\def\mbups{m_b^{1S}}

\def\vereq#1#2{\lower3pt\vbox{\baselineskip1pt\lineskip1pt
     \ialign{\\$#1\hfill##\hfil\\$\crcr#2\crcr\sim\crcr}}}

\newcommand{\Aerr}[2]{\ensuremath{\vphantom{0}^{+#1}_{-#2}}}
\newcommand{\lesssim}{\raisebox{-.25em}{$\stackrel{\normalsize <}{\sim}$}}

\begin{document}

\title[\sl 9th International Symposium on Heavy Flavor Physics]%
{Present and Future in Semileptonic $B$ Decays}

\classification{classification}
\keywords{keywords}

\author{Christian W.~Bauer
\thanks{Invited plenary talk presented at 9th International Symposium on Heavy Flavor Physics}
}
{address={Physics Department, University of California at San
Diego, La Jolla, CA 92093 }}

\copyrightyear  {2001}

\begin{abstract}
In this talk I review the status of our ability to extract the CKM matrix elements $|V_{ub}|$ and $|V_{cb}|$ from semileptonic decays. I will review both exclusive and inclusive methods and put a strong emphasis on how to ensure keeping the extractions model independent.
\end{abstract}

\date{\today}

\maketitle

\section{Introduction}
One of the main goals in particle physics over the next several years is to test precisely the flavor sector of the standard model (SM). In the SM, all weak interactions are determined by 12 parameters, namely the Fermi coupling constant, the weak mixing angle $\theta_W$, the six quark masses and four parameters in the CKM matrix. $B$ physics plays an integral role in overconstraining the four parameters in the CKM matrix, which leads to a stringent test of the SM. In this talk I will give an overview of the progress that has been made on using semileptonic $B$ decays to determine two of these parameters, the magnitude of $V_{ub}$ and $V_{cb}$. Semileptonic decays provide an ideal way to measure the magnitude of CKM matrix elements, since the strong interaction effects are greatly simplified by the presence of the two leptons in the final state. 

I will review how to measure $V_{ub}$ and $V_{cb}$ from both inclusive and exclusive decays. I will present the current status of extracting these CKM parameters from semileptonic decays, and point out possible improvements in the coming years. 

\section{Determination of $V_{\rm \lowercase{cb}}$}

\subsection{Exclusive Decays}
One of the main applications of heavy quark effective theory (HQET) \cite{IW} is that $V_{\rm cb}$ can be extracted in a model independent way using exclusive $b \to c$ semileptonic decays. Heavy quark spin and flavor symmetry predicts that in the infinite mass limit all heavy to heavy form factors are determined by a single nonperturbative function, the Isgur-Wise function. Moreover, this function is normalized to unity at zero recoil, which allows for a model independent prediction of semileptonic exclusive decays at that kinematic point. 

To be more precise, consider the decay $B \to D^* \ell \bar \nu$, and define the four-velocities of the $B$ and the $D^*$ mesons
\begin{eqnarray}
v_b^\mu = \frac{p_B^\mu}{m_B}\,, \qquad  v_c^\mu = \frac{p_{D^*}^\mu}{m_{D^*}}\,.
\end{eqnarray}
The decay rate as a function of the recoil $w = v_b \cdot v_c$ is given by
\begin{eqnarray}
\frac{d \Gamma_{B \to D^*}}{d w} = {\sqrt{w^2-1}} f(m_B,m_D,w) {|V_{cb}|^2} {F_{D^*}^2(w)}\,,
\end{eqnarray}
where $\sqrt{w^2-1}$ is a phase space suppression factor, $F_{D^*}^2(w)$ is a form factor and 
\begin{eqnarray}
f(m_B,m_D,w) = \frac{G_F^2m_B^5}{48 \pi^3} r^3(1-r)^2 (1+w^2)\left(1+\frac{4w}{1+w}\frac{1-2w r+r^2}{(1-r)^2} \right)\,,
\end{eqnarray}
where $r=m_{D^*}/m_B$. 
The form factor is related to the Isgur Wise function by 
\begin{eqnarray}
{F_{D^*}(w) = \eta_A \, \xi(w)}\,,
\end{eqnarray}
where $ \eta_A$ is a correction factor including both perturbative corrections, given  by an expansion in $\alpha_s(m_b)$, and nonperturbative corrections, determined by an expansion in $\Lambda/m_{b,c}$. At zero recoil the Isgur-Wise function is normalized, $\xi(1) =1$, and all ${\cal O}(\Lambda/m)$ corrections in $\eta_A$ vanish at that kinematic point \cite{lukestheorem}. The perturbative contributions to $\eta_A$ have been calculated to two loop order \cite{Czarnecki} and the $\Lambda^2/m_{(b,c)}^2$ corrections can be estimated using phenomenological models \cite{Falk_Neubert}. This leads to \cite{babarbook}
\begin{eqnarray}\label{FDmodel}
F_{D^*}(1) = \eta_A = 0.913 \pm 0.042\,,
\end{eqnarray}
where the uncertainty is mostly due to the unknown $\Lambda^2/m^2$ and higher corrections. 
Recently, this form factor has also been calculated on the lattice, and the result has identical central value and comparable uncertainties\footnote{This result does not include the QED corrections of $+0.007$, which is included in (\ref{FDmodel}).} \cite{latticeFD}
\begin{eqnarray}
F^{\rm Lat}_{D^*}(1) = 0.913 \Aerr{24}{17}\Aerr{17}{30}\,.
\end{eqnarray}

\begin{figure}
 \makebox{\includegraphics[height=.2\textheight]{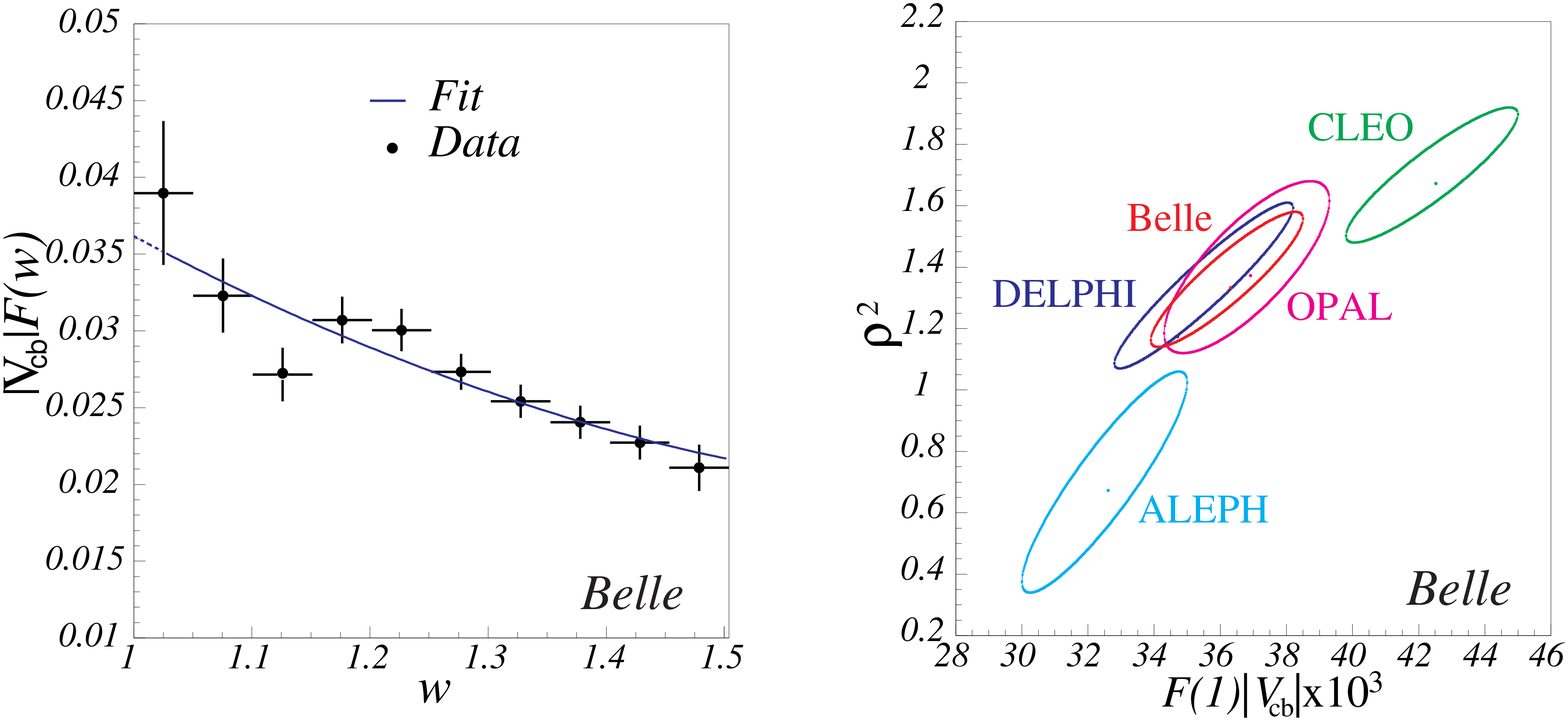}}
  \caption{Left: Extrapolation of $|V_{cb}| F_{D^*}(w)$ to the zero recoil point by Belle. Right: comparison of results from various collaborations.\label{vcbcombined}}
\end{figure}

Several experiments have measured this differential decay rate and for illustration we show in Fig.~\ref{vcbcombined} a plot of the BELLE collaboration together with a plot comparing the different measurements. 
Since the rate vanishes at zero recoil due to the phase space suppression factor, the differential rate has to be extrapolated to zero recoil. This extrapolation is guided by a model independent relationship between the curvature and the slope of $F_{D^*}(1)$ \cite{dispersion}. 
Combining the experimental measurement with the theoretical calculation of $F_{D^*}(1)$ leads to the values presented at this conference \cite{BelleVcb,CleoVcb,LEPVcb}
\begin{eqnarray}
V^{\rm BELLE}_{cb} &=& [39.7 \pm 1.6 (stat) \pm 2.0 (sys) \pm 1.9 (theor)] \times 10^{-3}\nonumber\\
 V^{\rm CLEO}_{cb} &=& [46.2 \pm 1.4 (stat) \pm 2.0 (sys) \pm 2.1 (theor)] \times 10^{-3}\\
 V^{\rm LEP}_{cb} &=& [40.5 \pm 0.5 \pm 2.0] \times 10^{-3}\\
\end{eqnarray}

 I would like to make a few comments on how the accuracy in this measurement might be improved. As can be seen from Fig.~\ref{vcbcombined}, there is a large correlation between the slope of the physical form factor and its value at zero recoil. Thus, information about this slope can reduce the errors on $|V_{cb}|$ considerably. There have recently been suggestions that QCD sumrules give constraints on the slope of the Isgur-Wise function \cite{Uraltsev}, although one has to be careful in relating these to the slopes of the physical form factor. Since the slopes of the $B \to D$ and the $B \to D^*$ form factor are related to one another in a calculable way, one should use the information from both of these decays at the same time (note that the experimental uncertainty in the slopes in these two decays is roughly the same \cite{BelleVcb}). In summary, performing a new complete high statistics analysis, including both the $D$ and $D^*$, as well as the form factor relations $R_1$ and $R_2$, should lower the uncertainties below the current level of 5\%.

\subsection{Inclusive Decays}

Inclusive semileptonic $B$ decays can be calculated using an operator product expansion (OPE). This leads to a simultaneous expansion in powers of the strong coupling constant $\alpha_s(m_b)$  and inverse powers of the heavy $b$ quark mass. At leading order in this expansion this reproduces the parton model result
\begin{eqnarray}
\Gamma_0 = \frac{G_F^2 |V_{cb}|^2 m_b^5}{192 \pi^3} \left( 1-8 \rho + 8 \rho^3 - \rho^4 -12 \rho^2 \log \rho \right)\,,
\end{eqnarray}
where $\rho = m_c^2/m_b^2$, and nonperturbative corrections are suppressed by at least two powers of $m_b$. The resulting expression for the total rate of the semileptonic $B \to X_c \ell \bar\nu$ can be written as
\begin{eqnarray}\label{Gammaincl}
\Gamma^{b \to c} = \Gamma_0 \left[ 1+ A \left[\frac{\alpha_s}{\pi}\right] + B \left[\left(\frac{\alpha_s}{\pi}\right)^2 \beta_0\right] + 0 \left[\frac{\Lambda}{m_b}\right] + C \left[\frac{\Lambda^2}{m_b^2}\right] + {\cal O} \left(\alpha_s^2, \frac{\Lambda^3}{m_b^3}, \frac{\alpha_s}{m_b^2} \right) \right]\,,
\end{eqnarray}
where the three coefficients $A,\, B, \, C$ depend on the quark masses $m_{(c,b)}$. The perturbative corrections are known up to order $\alpha_s^2\beta_0$. There are no  nonperturbative corrections at order $\Lambda/m_b$, and at order  $\Lambda^2/m_b^2$ they are given in terms of the two matrix elements $\lambda_1$ and $\lambda_2$, with the dependence on these matrix elements contained in the coefficient $C \equiv C(\lambda_1, \lambda_2)$. Since the value of $\lambda_2$ can be obtained from the $B - B^*$ mass splitting, and the charm quark mass can be related to known meson masses, the $b$ quark mass and $\lambda_1$, the only unknowns in extracting $|V_{cb}|$ from the inclusive rate are the precise value of the $b$ quark mass $m_b$ and the matrix element $\lambda_1$. 

In my opinion the best way to extract $m_b$ and $\lambda_1$ is to use the semileptonic decay data itself. Various differential decay spectra can be measured, not only the total decay rate. Several observables have already been constructed out of these spectra which are sensitive to $m_b$ and $\lambda_1$ \cite{moments}. The goal should be to measure and calculate as many inclusive observables as possible. All these observables are given in terms of $m_b$ and $\lambda_1$, so a fit to these observables will determine these parameters with theoretical uncertainties given by $\Lambda^3/m_b^3$ terms in the OPE \cite{mcubed}. Recently, the CLEO collaboration performed such an analysis in which they used the first moment of the hadronic invariant mass spectrum in $B \to X_c \ell \bar \nu$ and the first moment of the photon energy in the rare decay $B \to X_s \gamma$ to extract $m_b$ and $\lambda_1$ and used these values to determine $V_{cb}$ \cite{CleoVcb}. Defining $\bar\Lambda = m_B - m_b + \ldots$ they find
\begin{eqnarray}
\bar\Lambda = 0.35 \pm 0.07 \pm 0.10 \,\,{\rm GeV}\,, \qquad \lambda_1 = -0.238 \pm 0.071 \pm 0.078 \,\,{\rm GeV}^2\,,
\end{eqnarray}
and with that 
\begin{eqnarray}\label{vcbincl}
|V_{cb}| = [4.04 \pm 0.09 \pm 0.05 \pm 0.08 ] \times 10^{-3}\,.
\end{eqnarray}
The errors are (in order) from the measurement of the total decay rate, from the uncertainties in $\bar\Lambda$, $\lambda_1$, and from the higher order terms in both the perturbative and nonperturbative expansion. 

Using $\bar\Lambda$ as the mass parameter in the semileptonic decay rate has the disadvantage that the parameter $\bar\Lambda$ is not an infrared safe quantity. This means that the value of this parameter is ambiguous up to terms of order $\Lambda_{\rm QCD}$ \cite{Renormalon1}. This ambiguity cancels in any physical observable by a similar ambiguity in the perturbative expansion \cite{Renormalon2}.  For this cancellation to occur, however, it is essential that one works consistently to a particular order in perturbation theory. Since this is an artifact of the pole mass, it would be convenient to repeat the analysis with an infrared safe definition of the $b$ quark mass \cite{shortmass}. The numerical value of such a short distance mass can then be used safely in other processes as well.

Finally, I want to comment on an uncertainty not included in (\ref{vcbincl}), namely the uncertainty from duality violations, which are present for any inclusive observable in heavy quark decays. While there have been several estimates of these uncertainties \cite{duality}, I again would like to advocate using the semileptonic decay data itself to test for duality violations. Since many semileptonic observables can be calculated to the same order as (\ref{Gammaincl}) and all depend on the same two parameters $m_b$ and $\lambda_1$, one can check for duality violations by comparing the theoretical predictions with experimental results. This is an important analysis which still needs to be performed.

\section{Determination of $V_{\rm \lowercase{ub}}$}

\subsection{Exclusive Decays}

Measuring $|V_{ub}|$ from exclusive decays of $B$ mesons is considerably more difficult  than the determination of $|V_{cb}|$ discussed earlier. This is because the heavy quark spin-flavor symmetry, which gives rise to the simplifications of heavy to heavy form factors, does not apply to the heavy to light decays mediated by a $b \to u$ transition. Thus, to extract $|V_{ub}|$ from exclusive decays one has to find other ways to calculate   these form factors. 

As an example, consider the exclusive decay $B \to \pi \ell \bar\nu$. The required matrix element can be written in terms of two form factors
\begin{eqnarray}
\langle \pi | \bar{u} \gamma^\mu P_L | B\rangle = f_+(q^2) \left[ (p_B+p_\pi)^\mu - \frac{m_B^2}{q^2}q^\mu \right] + f_0(q^2) \frac{m_B^2}{q^2} q^\mu\,,
\end{eqnarray}
and a similar relation is true for $B \to \rho \ell \bar\nu$. 
The form factor $f_0(q^2)$ vanishes for zero lepton mass, but the form factor $f_+(q^2)$ is required to describe the decay rate. 
There are various suggestions in the literature how to extract these form factors. For $B \to \rho \ell \bar\nu$ the relevant form factors can be related to those in $D \to K^* \ell \nu$ using heavy quark and chiral symmetries\cite{Vubexcl1}. Corrections are first order in both heavy quark and chiral symmetry breaking individually and are at the 30\% level. This uncertainty could be reduced considerably if information from $D \to \rho \ell \nu$ and $B \to K^* \ell^+ \ell^-$ became available, since then a double ratio can be constructed which has corrections which simultaneously violate heavy quark and chiral symmetry \cite{Vubexcl2}.  The resulting uncertainties should be below the 10\% level. 

\begin{figure}
 \makebox{\includegraphics[width=6cm]{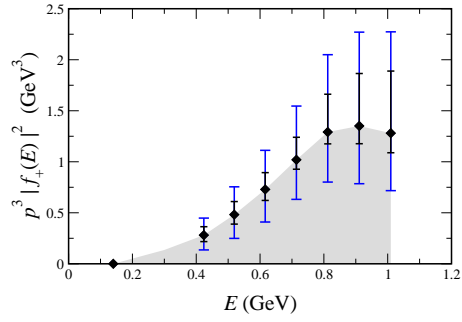}}
  \caption{Lattice calculation of the form factor $f_+$ as a function of the pion energy. \label{lattice}}
\end{figure}

Ultimately, lattice calculations will determine the required heavy to light form factors in a model independent way. In the last year several quenched calculation of the form factor $f_+(q^2)$ have been performed. Since lattice calculations can only be performed for slow pions, the lattice results are only available for small pion energies $E_\pi \lesssim 1\, {\rm GeV}$. A result from a recent lattice calculation of this form factor \cite{latticeVub} is shown in Fig.~\ref{lattice}. Experimentally, the situation is reversed with the efficiency to measure slow pions being small. In order to use lattice calculations to determine $|V_{ub}|$ from exclusive decays, however, it is crucial to increase the kinematic overlap of the lattice calculation of the form factors with the experimental measurements of the differential decay rates. 

\subsection{Inclusive Decays}
The inclusive decay rate $B \to X_u \ell \bar \nu$ is directly proportional to $|V_{ub}|^2$ and can be calculated reliably and with small uncertainties using the operator product expansion. Unfortunately, the $\sim$100 times background from $B \to X_c \ell \bar \nu$ makes the measurement of the totally inclusive rate an almost impossible task. Several cuts have been proposed in order to reject the $b \to c$ background, however care has to be taken to ensure that the decay rate in the restricted region of phase space can still be predicted reliably theoretically. The cut which is easiest to implement experimentally is on the energy of the charged lepton, requiring $E_\ell > (m_B^2-m_D^2)/2m_B$. Unfortunately, this cut restricts the remaining region of phase space too much for the OPE to still be valid. Instead, a twist expansion has to be performed \cite{shape}, and at leading order the decay rate is determined by the light cone distribution function of the $B$ meson, with subleading twist corrections suppressed by powers of $\Lambda/m_b$ \cite{BLM}. This distribution function is the matrix element of  a nonlocal operator and can not be calculated perturbatively. However, since it is a property of the $B$ meson itself, and thus independent of the decay process, it can be extracted from a different process and then used in the inclusive $b \to u \ell \bar\nu$ decay. This is similar to parton distribution functions, which can be measured for example in $e \, p$ collisions and then be used in other processes. The best way to measure this distribution function is in the decay $B \to X_s \gamma$. At leading order in $\alpha_s$, the shape of the photon energy spectrum is entirely due to this light cone distribution function, and the perturbative corrections are known to order $\alpha_s^2 \beta_0$ \cite{LLMW}. The photon spectrum has recently been measured to good accuracy by the CLEO collaboration \cite{CleoVcb}, which constituted the first determination of this structure function. 

Unfortunately, a cut on the lepton energy has the disadvantage that only $\sim 10 \%$ of the events pass this cut. Thus, the decay is far from being fully inclusive and one has to worry about duality violations becoming large. It is therefore advantageous to find ways to suppress the background which allows more $b \to u$ events to survive. The optimal such cut is on the hadronic invariant mass $m_X < m_D$ \cite{sHcut}. This cut is optimal in the sense that any other cut is a subset of it, and it has been estimated that $\sim 80\%$ of the $b \to u$ events survive. The same structure function which describes the lepton energy endpoint is needed to describe the region of low invariant mass. In \cite{mXrothstein} it was shown how to eliminate the structure function entirely and measure $|V_{ub}|$ model independently by using semileptonic data in combination with the photon energy spectrum in $B \to X_s \gamma$. The main uncertainties to this method of measuring $|V_{ub}|$ are given by unknown subleading structure functions, which are parametrically suppressed by $\Lambda_{\rm QCD}/m_b$ \cite{BLM}. It is hard to quantify the exact size of these uncertainties, and I would conservatively estimate them to be 10-20\%. 

It was shown in \cite{qsqcut} that using a cut on the leptonic invariant mass $q^2 = (p_\ell +p_\nu)^2$ allows to measure  $|V_{ub}|^2$ without requiring knowledge of the structure function of the $B$ meson. The number of events surviving such a cut on $q^2$ can be calculated using the usual local OPE and depending on the exact value of the cut chosen, the fraction of events surviving the cut is 10-20\%, with uncertainties on $|V_{ub}|$ ranging from 15\% for $q^2_{\rm cut} = m_B^2-m_D^2 = 11.6 \,{\rm GeV}^2$ to 25\% for $q^2_{\rm cut} = 14 \,{\rm GeV}^2$. The advantage of not having to use information from $B \to X_s \gamma$ and of the absence of power corrections at order $\Lambda_{\rm QCD}/m_b$ is offset by the fact that this cut again eliminates a huge fraction of the events and the worries about duality violations reappear. 

\begin{figure}
 \makebox{\includegraphics[width=10cm]{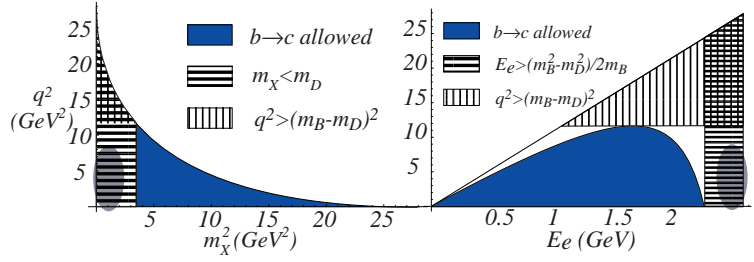}}
  \caption{The dalitz plot in the  $q^2/s_H$ and $q^2/E_\ell$ plane. The grey blob in the lower right (left) hand corner of the phase space indicates where the local OPE breaks down.\label{dalitzfigqsq}}
\end{figure}

To improve the situation, one has to understand why the $q^2$ cut allows to calculate using the local OPE, while the other cuts require the twist expansion. The twist expansion is relevant if the remaining phase space is dominated by hadronic states with energy much larger than their invariant mass. This is the case for both a cut on $E_\ell$ and a cut on $m_X$. The cut on $q^2$, however, restricts the hadron energy to be $E_X < m_B - \sqrt{q^2_c} < m_D$ and this cut therefore eliminates this dangerous region of phase space. The situation is illustrated in Fig.~\ref{dalitzfigqsq}. 

\begin{figure}
 \makebox{\includegraphics[width=10cm]{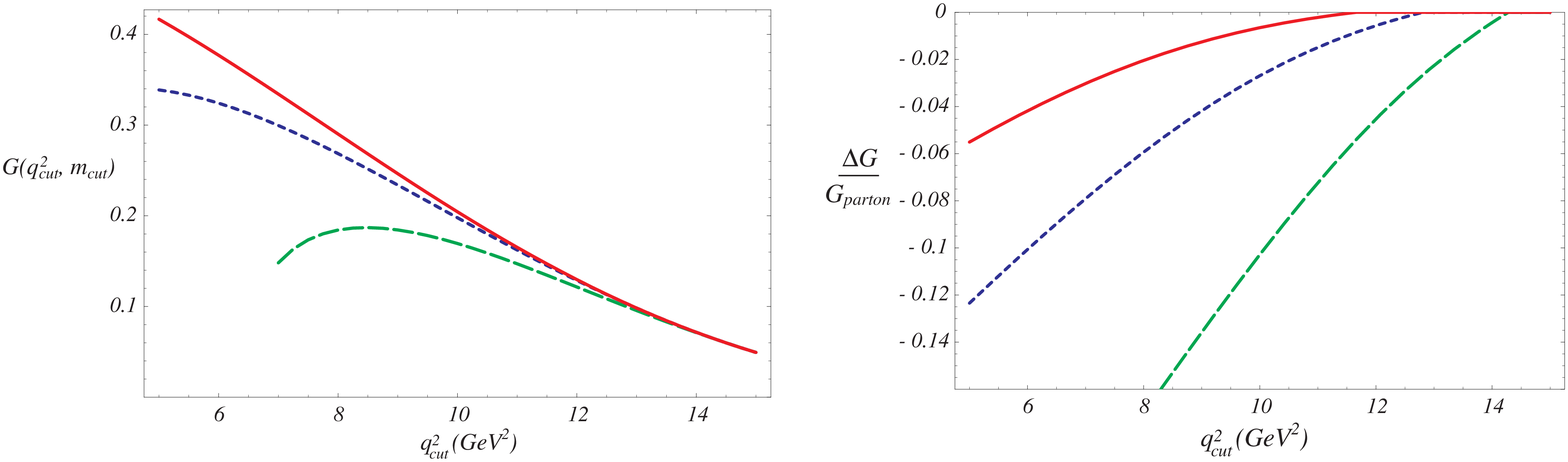}}
  \caption{Left: $\gcut$ as a function of $\qcut^2$, for $\mxcut=1.86\,\gev$ (solid  line), $1.7\,\gev$ (short dashed line) and $1.5\,\gev$ (long dashed line). Right: The effect of the structure function. \label{final_Sfig}}
\end{figure}

In \cite{doublecut} a new strategy to measure $|V_{ub}|$ was proposed which combines the advantages of the $m_X$ cut with those of the $q^2$ cut. The idea is to use a cut on the hadronic invariant mass to reject the $b \to c$ background, and simultaneously use a lower cut on $q^2$ to avoid the twist region. The pure $m_X$ and $q^2$ cuts are contained in this approach as the limits $q_{\rm cut}^2 = 0$ or $q_{\rm cut}^2 = (m_B - m_X^{\rm cut})^2$, respectively. Thus, varying the $q_{\rm cut}^2$ in the presence of a cut on $m_X$ allows to smoothly interpolate between these two cases. 
Using these results the strategy to extract $|V_{ub}|$ in a model independent way is: 
\begin{itemize}
\item make the cut on $m_X$ as large as possible, keeping the background from 
$b\to c$ small.
\item for a given cut on $m_X$, reduce the $q^2$ such as to minimize the overall uncertainty.
\end{itemize}
To illustrate this procedure I will present graphically the results of \cite{doublecut}. In Fig.~\ref{final_Sfig} the fraction of events surviving such combined cuts are shown for three values of $m_X^{\rm cut}$, in combination with the effect of the structure function. As advertised, lowering the cut on $q^2$ increases the fraction of events, while the effect of the structure function grows. Depending on how well the structure function will be known in the future determines how much the cut on $q^2$ can be lowered. Due to the fact that the structure function can only be determined up to corrections scaling as $\Lambda_{\rm QCD}/m_b$, it will always have an uncertainty at the 10-20\% level, even with a perfect measurement of the $B \to X_s\gamma$ photon spectrum. Thus, for a determination of $|V_{ub}|$ with uncertainties below 10\% one should ensure that the difference between the prediction with and without the structure function is less than $\sim 30 \%$. 

As in any inclusive observable, the remaining uncertainties are due to three sources: unknown perturbative corrections, uncertainties due to the parameters $m_b$ and $\lambda_1$ and uncertainties due to unknown matrix  elements of local operators at $O(\Lambda^3/m_b^3)$ in the OPE. To compare various methods to determine $|V_{ub}|$ it is crucial to have a consistent scheme to determine the theoretical uncertainties. In \cite{doublecut} such an error analysis was performed for the combined $q^2$, $m_X$ cut method, and in Table.~\ref{finaltable} I present the result for various different combinations of cuts. 
\begin{table}[ht]
\begin{tabular}{c||c|c|ccc|c}
Cuts on $(q^2,\,m_X^2)$  &  ~$Fract$~
   &  ~$\Delta_{\rm struct}V_{ub}$~ &  ~$\Delta_{\rm pert}V_{ub}$  &  
   $\matrix{\Delta_{m_b}V_{ub} \cr {\footnotesize \pm80/30\,\mev}}$  &
   $\Delta_{1/m^3}V_{ub}$~   &  $\Delta V_{ub}$  \\ \hline\hline
\multicolumn{1}{c}{Combined cuts} & \multicolumn{6}{c}{} \\ \hline
$6\,\gev^2, 1.86\,\gev$	&  46\%  &  $ -2\%$  &4\%  &  7\%/2.5\%  &  3\%  &
8\%/5\%  \\
$8\,\gev^2, 1.7\,\gev$ 	&  33\% &  $-3\%$&  6\%  &  8\%/3\%  &  4\%  &
9\%/6\% \\ 
$ 11\,\gev^2, 1.5\,\gev$  & 18\% &  $-4\%$ &13\% & 9\%/3.5\% & 8\% &
14\%/11\% \\ \hline\hline
\multicolumn{1}{c}{Pure $q^2$ cuts} & \multicolumn{6}{c}{} \\ \hline
~$(m_B-m_D)^2, m_D$~ & 0.17\% & --\,--&15\% &19\%/7\% & 18\% & 
~315\%/12\%
\end{tabular}\vspace*{4pt}
\caption{Fraction of events for several different choices of $(\qcut^2,\mxcut)$, along
with the uncertainties on $|V_{ub}|$.  The two last lines correspond to pure $q^2$ cuts,
$\mxcut = m_B-\sqrt{q^2}$, and are included for comparison.  $\Delta_{\rm
struct}$ gives the fractional  effect of the structure function $f(k_+)$ in
a simple model; we do not include an uncertainty on
this in our error estimate. The overall uncertainty $\Delta |V_{ub}|$ is obtained
by combining the other uncertainties in quadrature.   The two values correspond
to $\Delta \mbups=\pm 80\,\mev$ and $\pm 30\,\mev$.}
\label{finaltable}
\end{table}
From this table it is obvious that even without any knowledge of the structure function, and using a realistic cut of $m_X^{\rm cut} = 1.7\, {\rm GeV}$, $|V_{ub}|$ can still be measured with uncertainties below the 10\% level by choosing $q^2_{\rm cut} \approx 8 {\rm GeV}^2$. 
With many different methods to measure $|V_{ub}|$, the question arises of which measurement to ultimately use.  In my opinion, this question will most likely be answered once a full analysis is performed and the theoretical errors are investigated in conjunction with the experimental uncertainties. Depending on the details of the experimental efficiencies, some combination  (most likely non-linear) of the three cuts on $E_\ell$, $m_X$ and $q^2$ will prove to be the most effective way to minimize the overall uncertainty. The cut on the hadronic invariant mass $m_X$ will most likely be the best way to eliminate the $b \to c$ background, the cut on the leptonic invariant mass $q^2$ will limit the effect of the nonperturbative structure function and the cut on the lepton energy $E_\ell$ can help with experimental efficiencies. A careful analysis, which includes a realistic structure function and conservative errors should be performed to find the ideal combination of cuts.

\section{Conclusions}

In this talk I reviewed the current status of determining the magnitude of the CKM matrix elements $|V_{ub}|$ and $|V_{cb}|$ from semileptonic $B$ meson decays. For exclusive decays I reviewed theory and experiment and pointed towards a few improvements possible in the coming years. For inclusive decays, I reviewed  how their decay spectra can be calculated (up to duality violations) using an operator product expansion, which at leading order reproduces the parton model results. Nonperturbative corrections are parameterized by matrix elements of local operators suppressed by powers of $m_b$. There are no corrections at ${\cal O}(\Lambda/m_b)$, and at second order in the inverse heavy quark mass there are two matrix elements, commonly labeled by $\lambda_1$ and $\lambda_2$. While the value of $\lambda_2$ can be obtained from the $B - B^*$ mass splitting, the value of $\lambda_1$ is currently not known very well. Thus, in order to calculate the inclusive semileptonic observables precisely, we need to obtain values for the $b$ quark mass, which enters the calculation, as well as the for the parameter $\lambda_1$.

For $B \to X_c \ell \bar \nu$, these parameters give rise to the dominant theoretical uncertainty in the determination of $|V_{cb}|$ and their precise determination is therefore crucial to lower this uncertainty. I have argued here that these parameters should be extracted from decay spectra itself, and then used to extract $|V_{cb}|$. I also put forward the idea that duality violations can be checked using semileptonic decay spectra.

To measure $|V_{ub}|$ from the inclusive decay $B \to X_u \ell \bar\nu$ one has to deal with the large background from $b \to c$ transitions. Imposing kinematic cuts to suppress this background tends to destroy the convergence of the OPE and for both a cut on the lepton energy and a cut on the hadronic invariant mass an incalculable structure function is required, with corrections suppressed by $\Lambda/m_b$. This function can be extracted from the photon energy spectrum in $B \to X_s \gamma$, thus a precise measurement of this spectrum is very important. A cut on the leptonic invariant mass $q^2$ can be used to lower the effect of this structure function and is therefore crucial to minimize theoretical uncertainties. Ultimately, a dedicated analysis which combines the experimental and theoretical uncertainties will decide which combination of cuts will allow for the most precise determination of $|V_{ub}|$. I anticipate that such a study will eventually allow a model independent measurement of $|V_{ub}|$ with uncertainties at the 5\% level. 

\begin{theacknowledgments}
First, I would like to thank the organizers for organizing such a pleasant meeting, and for dealing so well with the unforeseeable circumstances. I would also like to thank Zoltan Ligeti and Michael Luke for collaborations on some of the work presented here, and Thomas Mannel for comments on the manuscript. This work was supported by the Department of Energy, 
Contract DOE-FG03-97ER40546
\end{theacknowledgments}
\pagebreak

\end{document}